\documentstyle[editedvolume,psfig]{crckapb} 

\begin{opening}
\title{PROPERTIES OF GALAXIES IN AND AROUND VOIDS}
\author{ULRICH HOPP}
\institute{Universit\"atssternwarte M\"unchen\\
           Scheiner Str. 1, D 81679 M\"unchen\\
           email: hopp$@$usm.uni-muenchen.de}
\end{opening}

\runningtitle{Void galaxies}

\begin{document}

% The \begin{document} command comes after the \end{opening}
% command.

\section*{Abstract}

\small{
Two surveys for intrinsically faint galaxies towards nearby voids have
been conducted at the MPI f\"ur Astronomie, Heidelberg. One selected
targets from a new diameter limited ($\Phi \ge 5''$) catalog with
morphological criteria while the other used digitized objective prism
Schmidt plates to select mainly HII dwarf galaxies. For some 450
galaxies, redshifts and other optical data were obtained.  We studied
the spatial distribution of the sample objects, their luminosity
function, and their intrinsic properties.

Most of the galaxies belong to already well known sheets and
filaments. But we found about a dozen highly isolated galaxies in each
sample (nearest neighborhood distance $\ge 3 h_{75}^{-1} Mpc$). These
tend to populate additional structures and are not distributed
homogeneously throughout the voids. As our results on 'void galaxies'
still suffer from small sample statistics, I also tried to combine
similar existing surveys of nearby voids to get further hints on the
larger structure and on the luminosity function of the isolated
galaxies.  No differences in the luminosity function of sheet and void
galaxies could be found.

The optical and infrared properties of both samples are in the normal
range for samples dominated by late-type dwarfs. Follow-up HI studies
show that the isolated dwarfs in both samples have unusual high
amount of neutral gas for a given luminosity.}

\section{Introduction}

One of the surprising results of the first redshift surveys was the
detection of empty regions in space, the voids. Some theories
predicted that a homogeneously distributed population of dwarf
galaxies may exist inside these voids (Dekel \& Silk, 1986), others
predicts at least filaments of matter crossing the empty regions of
the nearby universe (see Colberg et al, these proceedings). In 1990,
we started a survey for intrinsically small and faint galaxies towards
nearby voids with the idea that some selection effects introduced by
the catalogs and observers might have prevented the detection of a
void population (Hopp \& Kuhn, 1995). Obviously, dwarf galaxies of all
types with their faint absolute magnitudes, often rather low surface
brightness (SB), and small or even vanishing apparent diameters on
Schmidt survey plates are easily missed in all kind of
surveys. Indeed, it is known that dwarfs dominate the universe in
number ($\ge 65\%$\footnote{At least in the 10 Mpc sample of
Kraan-Korteweg \& Tammann (Schmidt \& Boller, 1992, Karachentsev,
1994).}), but form only a minor contribution ($\le 10\%$) to the
catalogs ('Zwicky', UGC) on which the early redshift surveys like the
CfA (Huchra et al, 1990) and its southern counterpart (da Costa et
al., 1994) were based.

Early attempts to clarify whether or not dwarfs are distributed like
the giants in the standard catalogs failed as they were limited by
the standard catalogs (e.g. Thuan et al, 1987). They were not able
to detect real dwarf galaxies even in the distance of the nearest
voids (Binggeli et al 1990, Hopp 1994). But these surveys already
showed that one has to deal with really faint objects\footnote{E.g.:
$M_B \ge -16^m, m_B \ge 18^m, SB_0 \ge 22^m/\Box'', r_s \le 1 kpc \sim
3''$.}, and therefore has to study the nearby voids ($v_R \le 10^4 km
s^{-1}$). In this sense, the famous Bo\"otes void (Kirchner et al.,
1981) which attracted otherwise very important survey work
(e.g. Weistrop et al, 1995, Szomoru et al. 1996), is too distant. 
Indeed, most of the so-far identified $\sim 60$ galaxies in
the Bo\"otes void are not dwarfs.

Here, I will report about the two Heidelberg-void surveys. I
will especially describe the properties of the very isolated galaxies which we
identified in and around voids.  Meanwhile, other independent surveys
were done, mainly by searching emission line galaxies (ELG, see
Popescu et al, 1996, for references), or galaxies of low
surface brightness (LSBG, e.g. Bothun et al, 1986, Mo et al, 1994,
Roennback \& Bergvall, 1996, Bothun et al, 1997, Impey et al 1996).
Two surveys were dedicated to morphological selection
(Eder et al, 1989, Binggeli et al 1990). I will compare to these 
other surveys and show that a combination of the catalogs to a
common data base has great potential in studying the matter
distribution in and around voids.  All surveys together may serve as
an ideal data base to reconstruct the mass distribution from the
peculiar motion of the galaxies (following e.g. da Costa et al, 1996),
especially, when the gaps between them will be filled.  Throughout
this paper, I use $H_0 = 75 km s^{-1} Mpc^{-1}$.

\section{Two Heidelberg surveys}

The shape of the luminosity function of galaxies yields magnitude
limited galaxy catalogs which are dominated in number by $L^*$
galaxies (Dickey, 1988). Increasing the magnitude limit of a catalog
thus largely increases the number of $L^*$ galaxies at large distances
while it adds only a few to the nearer dwarfs. Therefore,
the strategy of a general redshift survey like the LCRS or the
SLOAN\footnote{For a description of these two major redshift surveys
see the contributions by the two teams in these proceedings.} was
neither efficient nor feasible for our purpose of finding dwarfs in
the redshift distance of the nearest voids. Similar arguments hold
for diameter limited catalogs (Binggeli et al, 1990). Selection
affects in SB can very powerful effect the results of
any survey (Disney, 1976, Bothun et al. 1997). Thus, we started
dedicated surveys towards a few selected nearby voids. We tried to
optimize the selection for dwarf galaxies and LSBG
in and around these voids (Hopp 1994, Hopp \& Kuhn 1995). The
data were reported in Hopp et al. (1995, paper 1 hereafter), and
Popescu et al. 1996 (paper 2) while the spatial distribution of the
samples in discussed in Kuhn et al (1997, paper 3), and Popescu et al
(1997, paper 4).

\subsection{Void selection}

When we started, few void catalogs were available. Therefore, we
used the available cone diagrams (mostly from CfA1), and
selected those voids which have a diameter of $\ge 20 Mpc$, and are
completely free of Zwicky and UGC galaxies. We selected four fields
where $\mid b \mid \ge 40^o$.  Our fields cover a total of 10 voids up
to $10^4 km s^{-1}$ (see paper 1, 2). Later, we learned that our
naively defined voids agree quit well with more objective
definitions (Kauffmann \& Fairall, 1991, Saunders et al 1991, Slezak et
al 1993, Lindner et al, 1995).

\subsection{Selection effects}

The redshift surveys available at that time (CfA, SSRS) were based on
traditional catalogs (Zwicky et al., 1961-65, Nilson, 1973, Lauberts
1982) which are limited either in apparent magnitude of B $\le 15.5^m$
or in diameters $\Phi \ge 1.0'$. They are most severely limited in SB
due to the POSS I or ESO quick B capabilities. A discussion of the
structural properties of very nearby dwarfs forces one to include much
smaller diameters and also objects of lower SB (Hopp, 1994; footnote
2).

We obtained deep Calar Alto 3.5m prime focus images towards the center
of three of our fields and establish a diameter-limited catalog down
to $\Phi \ge 5''$. In the surroundings of these deep images, I
surveyed the POSS down to $\Phi \ge 21''$. Figure 1 in paper 1 shows
that the resulting sample fits nicely to the diameter distribution of
the UGC, going to smaller values. From this calatalogue, we selected
dwarf candidates by their morphology or low SB. The follow-up redshift
survey revealed a high percentage ($\sim 70\%$) of ELG in this sample
(paper 1) and a higher-than-average rate of ELGs among the isolated
galaxies.  Therefore, we decided to start a second survey for ELG's in
the Hamburg Quasar Survey data base (Hagen et al, 1995) which is
decribed in detail by Popescu (these proceedings, paper 2). The ELG
survey is only seeing limited in $\Phi$, but galaxies with large
$\Phi$ in their emission regions are hard to detect on objective prism
plates. Both Heidelberg void surveys cover the range $15 \le B \le 20.5^m$
($ 20.0 \le \mu_B [m/\Box''] \le 24.5$). The magnitude completeness
limit of the morphological survey is $B \sim 19.0^m$. The leading
selection effect for the ELG sample is the equivalent width of the
emission line used for selection, in our case [OIII]5007. The limit is
0.8 nm (see Popescu, this volume, and paper 4).

\subsection{The morphological survey}

This survey covered three fields towards nearby voids. A total of
$\sim 175$ redshifts were obtained together with B and R CCD images of
most of the galaxies (paper 1). The SB(r) of the galaxies in a
subsample was obtained by Vennik et al (1996, paper 5). Half of the
galaxies have redshifts above the limits of the CfA which is the
bright comparison sample in our study and are therefore useless for
our goal.  The average surface density of all objects is about
$1/\Box^{o}$.  Cone diagrams of all fields and a detailed discussion
of the nearest neighbor distance $D_{NN}$ distribution are the
essentials of paper 3.

\subsection{The emission line galaxy survey}

The survey was obtained in four fields, some identical to the
fields of the morphological survey. A total of $\sim 250$ redshifts
were obtained. For most of the galaxies, we have R or B CCD images,
for some both colors, which are now under study. Most of the spectra 
are good enough to discuss the abundances of the objects (in prep.). 
The data were presented in paper 2 while the spatial distribution and
the details of the selection effects are discussed in paper 4 and by
Popescu in these proceedings. This survey was pretty successful in
finding dwarf galaxies below $v_r \le 10^4 km s^{-1}$ (see Fig. 2
of Hopp \& Kuhn 1995).

\section{Spatial distribution}

For a detailed discussion, including cone diagrams, we refer to papers
3, 4 as well as to Popescu in these proceedings.  We were able to
detect about 25 highly isolated galaxies. We call a galaxy highly
isolated when $D_{NN} \ge 3.0 h_{75}^{-1} 3 Mpc$ and $cz \le 10^4 km
s^{-1}$ (Fig. 1).\footnote{One should not forget that $D_{NN}$ can be
measured {\it only} to the next {\it catalogued} galaxy!} All other
galaxies ($\sim 95\%$) follow well known sheets and filaments or are
situated in the background where the sampling is too sparse for any
density estimates.

Many of the highly isolated galaxies tend to populate the rims of the
voids. This may simply reflect the fact that late type dwarfs have the
lowest clustering power and drop off with a shallower gradient from
the density peaks of the sheets. 

A few are located more to the central parts of the voids. These
galaxies show some tendency to align themself in very sparsely
populated chains or filaments which devide bigger voids into smaller
entities (paper 4).  Filaments which are populated only by low
luminosity galaxies are an observational argument in favour of
hierarchical structure formation (Vogeley et al 1996).  Unfortunately,
we have only very few cases at hand in our surveys, but similar
conclusion were based on other ELG surveys (Lindner et et 1996).  We
estimated the number density contrast between the sheet galaxies and
those few in the voids to be $\Delta \rho / \rho \le 0.1$ (paper 3,
4).

\subsection{Similar surveys}

\begin{figure}[ht]      
\psfig{figure=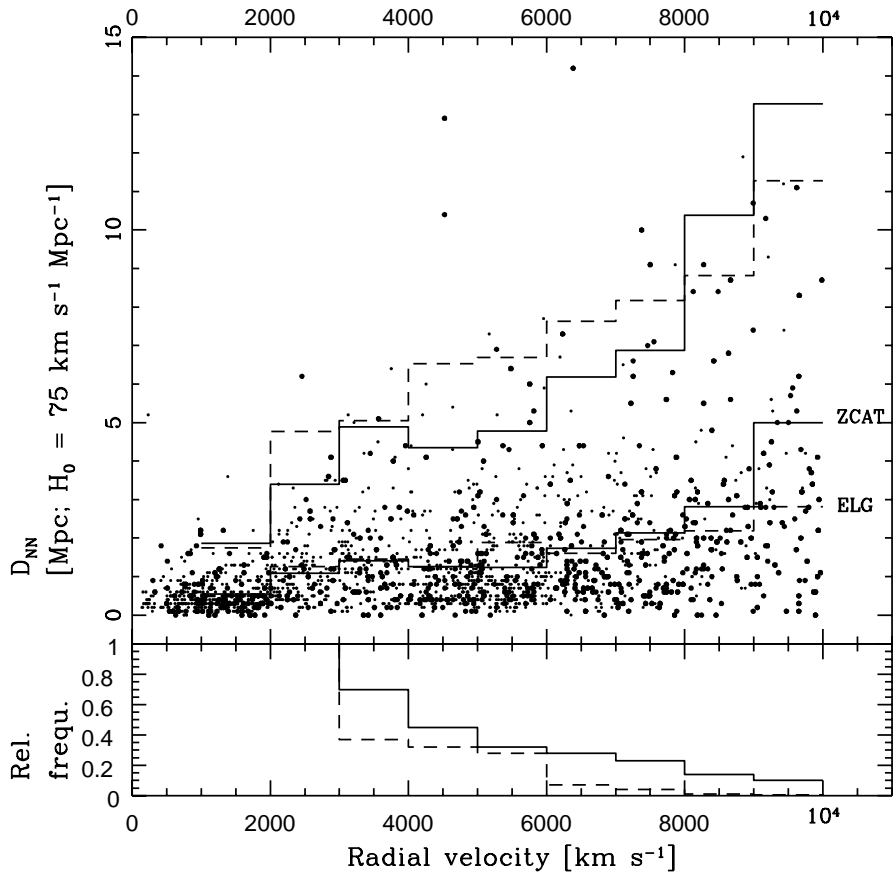,width=120mm,clip=t}
\caption{
{\it Top:} Nearest neighborhood distance $D_{NN}$ distribution as a function of
radial velocity for the MDS galaxies (big symbols) and the CfA
galaxies from the same volume (small). The lower lines indicate the
mean values in redshift bins of $10^3 km s^{-1}$ for the CfA (solid)
and the MDS (dashed), the upper the mean values plus $3 \sigma$. {\it
Bottom:} Relative frequency of galaxy numbers in MDS and CfA normalized
to the bin at $2500 km s^{-1}$. Obviously, the sampling decreases in
both samples with increasing redshift in a similar manner. Thus, the
absolute values of $D_{NN}$ at low and high redshifts can not be directly
compared. Therefore, we restricted the MDS to the interval $3..6 10^3
km s^{-1}$ for comparison of high and low density galaxies
with a cut between this two regimes at $3.0 h_{75}^{-1} Mpc$.}
\end{figure}

Having so few isolated galaxies in our samples, I tried to combine our
surveys with other surveys which are similar in selection functions,
covered field, volume, velocity and magnitude range. These are the SBS
survey (Pustil'nik et al 1995), the UM- and Case surveys by Salzer
(1989) and Rosenberg et al (1994), respectively. For a comparison of
these different surveys see paper 2. I call the merged dwarf sample
MDS. Naturally, this merging of catalogs yields an inhomogeneous
sample (Hopp, 1997). But this exercise was mainly done to convince
myself that a survey which will cover the whole nothern galactic cap
and link all the above mentioned surveys, including the two Heidelberg
void surveys, will be the right way to proceed. A project (under the
name Hamburg-SAO survey) was already started (Ugryumov et al 1997). We
obtained some further 75 redshift of ELGs which are already included
in the MDS.

From the same fields as the surveys, I extracted the CfA galaxies as a
comparison sample. In total, the MDS covers $4750 \Box^o$ and contains
1012 (and 877 CfA) galaxies within $v_R \le 10^4 km
s^{-1}$. All individual surveys show a few isolated galaxies as
discussed by their authors. 

Contrary to the authors of the ELG surveys, those who analysed the
spatial distribution of LSBG (e.g. Eder et al., 1989, Mo et al, 1994,
Schombert et al, 1997) found that while these LSBG are avoiding high
density regions, they else still follow the structures occupied by the
giants. As there is still some discrepancy in the
interpretation\footnote{This comes as a little surprise as the LSBG cone
diagrams shows also very isolated objects and statistical results are
similar. I was always left with the impression that the discrepancy is
more of semantic nature depending on the topic the authors like to
address. ELG and LSBG surveys both rule out the high-biasing scenario
that (irregular) dwarfs fill the voids.} , and as the selection
function surely differs from those of the ELG samples, I did not
include these surveys into the MDS so far.

\begin{figure}[ht]      
\psfig{figure=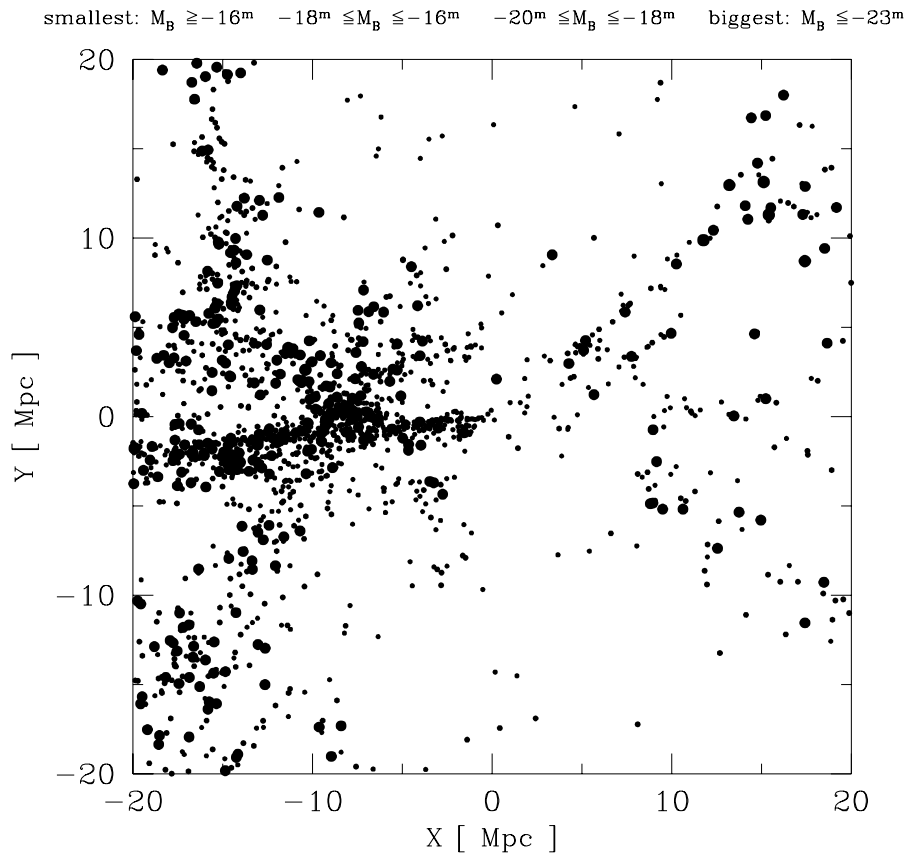,width=120mm,clip=t}
\caption{
Spatial distribution of the merged dwarf sample and the CfA galaxies
in a cube of $20 h_{75}^{-1}$ Mpc side length. Third coordinate is
projected. The symbol sizes indicates the {\it absolute} magnitude of
the galaxies. Besides the well-known rich structures which are
populated by galaxies of all luminosities, there exist regions which
are only populated by low-luminosity galaxies.}
\end{figure}

Most of the MDS galaxies follow the distribution outlined by the CfA
galaxies, namely the well-known structures which are defined by the 
massive galaxies. About 140 of the MDS galaxies are very isolated with 
$D_{NN} \ge 3.0 h_{75}^{-1} Mpc$ (Fig. 1). These isolated galaxies 
are not randomly distributed, but
they occupy the rims of the voids or form additional filaments which
seems to be populated only by low luminosity galaxies (Fig. 2; Hopp,
1997, paper 4). Similar results have been found by Lindner et al (1996)
and Vogeley et al (1994).

\section{Properties of isolated galaxies}

\subsection{Surface photometry}

As far as we finished the analysis, most of our objects are disk
galaxies with rather blue colors, quite common for a dwarf sample. The
mean scale length for the dwarfs is $2.1\pm0.8 h_{75}^{-1} kpc$ 
(paper 5). The isolated galaxies do not differ
in their structural properties. In the theoretical SB versus scale 
length diagram shown by Dalcanton during this conference, our sheet
and isolated galaxies distribute between the diameter limited UGC sample
and the Virgo cluster sample in the regime of relatively small scale
length and over nearly the whole range in SB (Dalcanton et al, 1997). Again,
no difference is obvious between isolated and sheet galaxies. Most
objects have central SB well below the 'Freeman disk'.

\subsection{Luminosity function}

I calculated the luminosity function for our sample as for the other
ELG samples (Fig 3) after applying $V/V_m$ test completeness
corrections for a volume limited sample. They all are pretty similar
and show a steep faint end slope. Then, I calculated the luminosity
function for the MDS within $3000 \le cz [km s^{-1}] \le 6000$ as the
completeness in this redshift range drops only slowly for dwarfs
(Fig. 1). The calculation was done separately for low and high density
regions (dividing line: $D_{NN} = 3.0 h_{75}^{-1} Mpc$).
Within the rather poor statistics for the low density regime ($\sim
140$ galaxies), the low and high density luminosity function are identical.

\begin{figure}[ht]      
\psfig{figure=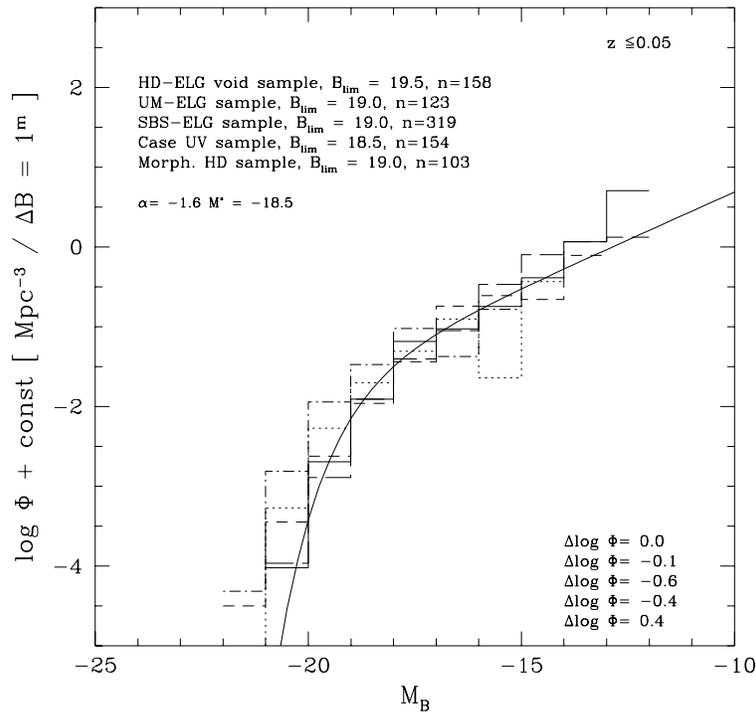,width=100mm,clip=t}
\caption{
Luminosity function derived from our surveys and several similar
sample from the literature (histogram). The line shows a
Schechter function with the indicated parameters for comparison, it is
not a fit. All these field dwarf samples tend to show a steep faint end slope.
The differences between the samples is negligible within the given
statistical and systematic errors.}
\end{figure}

\subsection{High HI abundances}

We obtained HI observations for subsamples of both Heidelberg void
surveys, mostly with the Effelsberg 100m antenna. The detection rate
was high ($\sim 67\%$ and $\ge 50\%$, respectively) and many galaxies
show a high amount of HI gas. The line profiles indicate in most cases
a dwarf galaxy. Especially, we found a trend that the HI-mass to blue
luminosity ratio of dwarfs for a given luminosity increases with
decreasing galaxy density, from the Virgo cluster to sheets to voids
(Huchtmeier et al 1997).  A first analysis of the ELG data also
supports this finding. No significant deviation from the Tully-Fisher
relation was found so far, including the void galaxies.

\subsection{Other frequencies}

With the aid of NED\footnote{The NASA/IPAC extragalactic database
(NED) is operated by the Jet Propulsion Laboratory, California
Institute of Technology, under contract with the National Aeronautics
and Space Administration.} and the recently publicly available ROSAT
point source catalog (Voges et al, 1996), we checked whether the
galaxies found in our two surveys were detected at other frequencies,
too. No galaxies from the morphological approach were detected by
IRAS, they do not shine up in the ROSAT point source catalog, nor does
NED list any radio continuum detection.

\begin{figure}[ht]      
\psfig{figure=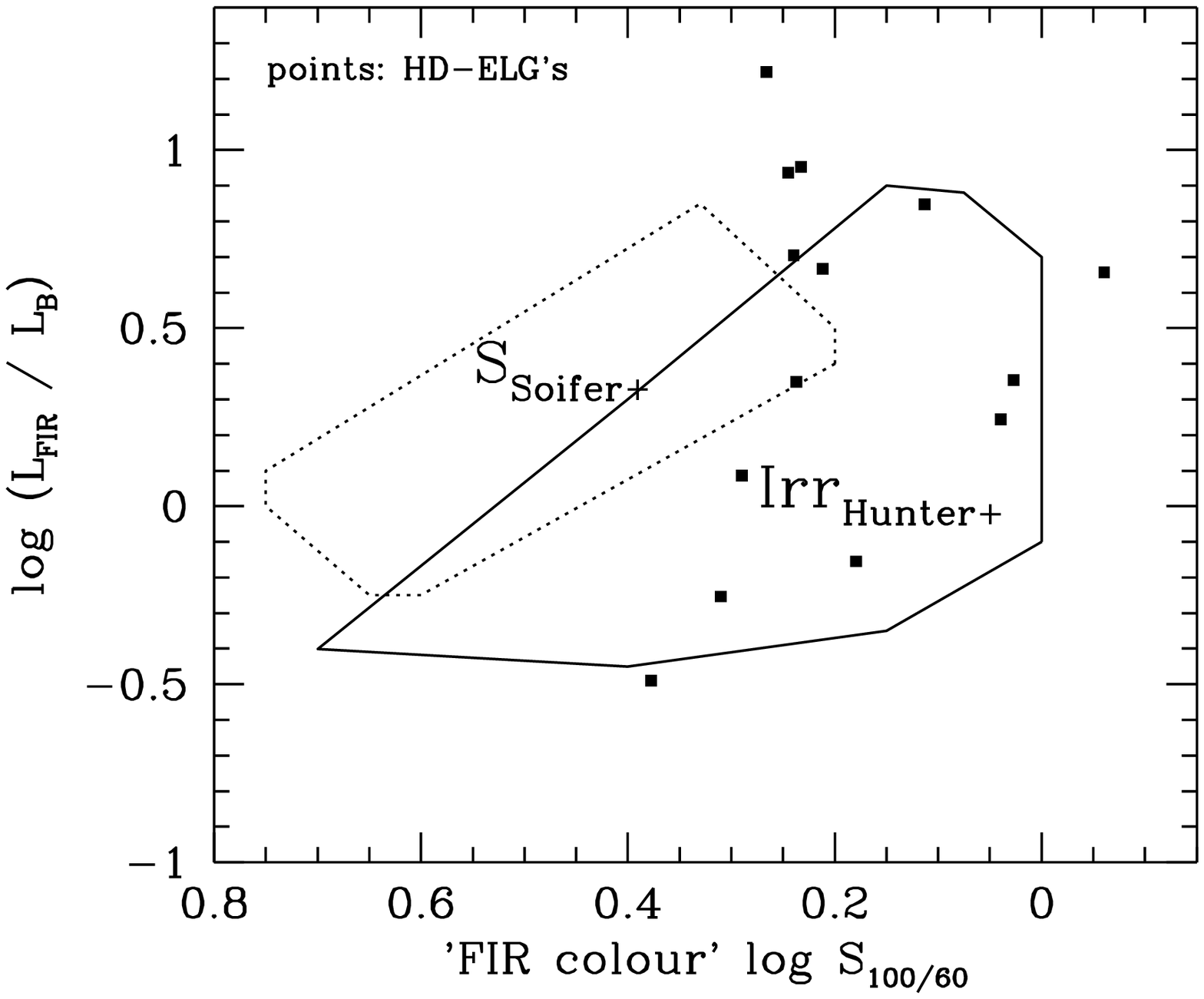,width=100mm,clip=t}
\caption{
Plot of FIR color index in 60 $\mu$ through 100 $\mu$ versus
ratio of FIR luminosity to B luminosity. Data from the IRAS point
source catalgue for those of the Heidelberg-ELG sample which were detected by
IRAS in both bands. The lines indicate the distribution regions for
spiral galaxies according to Soifner et al (1987) and for
irregular galaxies according to Hunter et al (1989). Most of our
galaxies occupy the region of irregular dwarfs.
}
\end{figure}

Contrary, about 15\% of the ELG galaxies were detected by IRAS in the
60 $\mu$ band, but none of the detected sources is highly isolated.
Some of them were also detected at 100 $\mu$ or at 25 $\mu$. In
Fig. 4, we show a FIR-B color-color plot where the detected objects
are indicated together with the general distribution for spirals and
irregular dwarfs. These detections again show that our ELGs occupy
mostly the dwarf parameter space. Only one single source may have been
detected by ROSAT. Seven 20 cm FIRST detections are available from NED.

\section{Conclusions}

I presented our two Heidelberg void surveys which each identified
about a dozen of highly isolated galaxies inside the voids. Some of
them are distributed along the rims of the voids while others
populate filaments across the voids. Some voids 
remain completely empty. Most of the galaxies are dwarfs with quite
normal optical properties, but high HI mass. Our results agree
well with the results of other surveys which detected dwarfs in and
around voids. From a combination of all ELG surveys we found no
evidence for a difference in the luminosity function of isolated and
sheet galaxies.  The number density contrast between sheets and the
voids remains high ($\Delta \rho / \rho \le 0.1$) despite the newly
identified void galaxies. 

\small{
I like to thank for many useful discussion Drs. R. Bender,
B. Binggeli, J. Einasto, H. Elsaesser, H.-J. Hagen, W.K. Huchtmeier, 
B. Kuhn, C.C. Popescu, S. Pulstil'nik, J. Roennback,
J. Salzer, and J. Vennik.  Especially, I remember with a deep and warm
feeling the discussions with the late Valentin Lipovetzky who died the
Sunday just before this meeting. I like to dedicate this paper to the
memory of Valentin.  It's a pleasure to thank Ulrich Thiele and the
Calar Alto crew for the support during the many observations.  The
author acknowledges the support of the SFB 375 of the Deutsche
Forschungsgemeinschaft.}

{\small \noindent {\bf References:}\\
%\begin{thebibliography}{} 
Binggeli,B., Tarenghi,M., Sandage,A., 1990, {\it A\&A} {\bf 228}, 42.\\
Bothun,C.D., Beers,T.C., Mould,J.R., Huchra,J.P., 1986, {\it ApJ} 
{\bf 308}, 510\\
Bothun,C.D., Impey,C., McGaugh,S., 1997 {\it PASP} in press\\
da Costa, L.N., Geller,M.J., Pellegrini,P.S., Latham,D.W.,
Fairall,A.P., Marzke,R.O., Willmer,C.N.A., Huchra,J.P.,
1994, {\it ApJ} {\bf 424}, L1\\
da Costa, L.N., Freudling,W., Wegner,G., Giovanelli,R., Haynes,M.P.,
Salzer,J.J., 1996, {\it ApJ} {\bf 468}, L5\\
Dalcanton,J., Spergel,D.N., Summers,F.J., 1997 {\it ApJ} preprint\\
Dekel,A., Silk,J., 1986, {\it ApJ} {\bf 303}, 39\\
Dickey,J.M., 1988, {\it Astron.Soc.Pac.Conf.Ser.} {\bf 5}, 9\\
Disney,M.J., 1976, {\it Nature} {\bf 263}, 573\\
%\bibitem{}
Eder,J.A., Schombert,J.M., Deckel,A., Oemler,A., 1989, {\it ApJ} 
{\bf 340}, 29\\
%\bibitem{}
Hagen,H.-J., Groote,D., Engels,D., Reimers,D., 1995, {\it A\&AS}
{\bf 111}, 195\\
%\bibitem{}
Hopp, U.: 1994, in: G.Meylan and P.Prugniel (eds.), {\it Dwarf Galaxies}
ESO Conf. Proceedings {\bf 49}, 37\\
%\bibitem{}
Hopp,U., 1997 {\it Proc. Symp. IAU} {\bf 179}, in press\\
%\bibitem{}
Hopp,U., Kuhn,B., Thiele,U., Birkle,K., Els\"asser,H., Kovachev,B.,
({\bf paper 1}) 1995 {\it A\&AS} {\bf 109}, 537 \\
%\bibitem{}
Hopp,U., Kuhn,B. 1995 {\it Reviews in Modern Astronomy}, {\bf 7}, 277\\
%\bibitem{}
Huchra,J.P., Geller,M.J., de Lapparent,V., Corwin,H.G. 1990) {\it ApJS} {\bf 72}, 433\\
%\bibitem{}
Huchtmeier,W.K., Hopp,U., Kuhn,B. 1997 {\it A\&A} accepted \\
%\bibitem{}
Hunter,D., Gallagher,J.S., Rice,W.L., Gillett,F.C., 1989, {\it ApJ}
{\bf 336}, 152\\
%\bibitem{}
Impey,C.D., Sparberry,D., Irwin,M.J., Bothun,G.D., 1996, {\it ApJS}
{\bf 105}, 209\\
%\bibitem{}
Karachentsev, I.D., 1994, {\it A\&Ap Transactions} {\bf 6}, 1\\
%\bibitem{}
Kauffmann,G., Fairall,A.P., 1991 {\it MNRAS} {\bf 248}, 313\\
%\bibitem{}
Kirchner,R.P., Oemler,A., Schechter,P.L., Schectman,S.A.: 1981, {\it ApJ}
{\bf 248}, L57\\
%\bibitem{}
Kuhn,B., Hopp,U., Els\"asser,H. ({\bf paper 3}, 1997 {\it A\&A} {\bf
318}, 405\\
%\bibitem{}
Lauberts,A., 1982, {\it The ESO/Uppsala Survey of the ESO (B) Atlas},
ESO, M\"unchen.\\
%\bibitem{}
Lindner,U., Einasto,M., Einasto,J., Freudling,W., Fricke,K., Lipovetsky,V.,
Pustilnik,S., Izotov,Y., Richter,G., 1996 {\it A\&A} {\bf 314}, 1\\
%\bibitem{}
Mo,H.J., McGaugh,S., Bothun,G.D., 1994 {\it MNRAS} 267, 129 \\
%\bibitem{}
Nilson,P., 1973, {\it Uppsala General Catalogue of Galaxies}, Nova 
Acta Reginae Soc. Sci. Uppsalinisis Ser. V: A Vol. {\bf 1} (UGC).\\
%\bibitem{}
Popescu,C.C., Hopp,U., Hagen,H.-J., Elsaesser,H., ({\bf paper 2},
1996 {\it A\&AS} {\bf 116}, 1\\
%\bibitem{}
Popescu,C.C., Hopp,U., Elsaesser,H., ({\bf paper 4}, 1997 {\it A\&A} submitted\\
%\bibitem{}
Pustil'nik,S.A., Ugryumov,A.V., Lipovetsky, A.A., Thuan,T.X.,
Guseva,N.G., 1995 {\it ApJ} {\bf 443}, 499 \\
%\bibitem{}
Roennback,J., Bergvall,N., 1996, {\it A\&A} {\bf 302}, 353\\
%\bibitem{}
Rosenberg,J.L. et al. 1994 {\it AJ} {\bf 108}, 1557 \\
%\bibitem{}
Salzer,J.J. 1989 {\it ApJ} {\bf 347}, 152 \\
%\bibitem{}
Saunders,W., Frenk,C., Rowan-Robinsom,M., Efstathiou,G., Lawrence,A.,
Ellis,R., Crawford,J., Xia,X.Y., Parry,I., 1991, {\it Nature} {\bf 349}, 32\\
%\bibitem{}
Schmidt, K.H., Boller, T., 1992, {\it AN} {\bf 313}, 329\\
%\bibitem{}
Schombert,J.M., Pildes,R.A., Eder,J.A., 1997, {\it ApJS} accepted\\
%\bibitem{}
Slezak,E., de Lapparent,V., Bijaoui,A., 1993, {\it ApJ} {\bf 409}, 517\\
%\bibitem{}
Soifer,B.T., Houck,J.R., Neugebauer,G., 1987, {\it ARA\&A} {\bf 25}, 187\\
%\bibitem{}
Szomoru,A., van Gorkom,J.H., Gregg,M.D., Strauss,M.A., 1996, {\it AJ}
{\bf 111}, 2150\\
%\bibitem{}
Thuan,T.X., Gott,J.R., Schneider,S.E., 1987, {\it ApJ} {\bf 315}, L93\\
%\bibitem{}
Vennik,J., Hopp,U., Kovachev,B., Kuhn,B., Elsaesser,H., {\bf paper
5}, 1996 {\it A\&AS} {\bf 117}, 261 \\
%\bibitem{}
Vogeley,M.S., Geller,M.J., Park,C., Huchra,J.P., 1994, {\it AJ} 
{\bf 108}, 745\\
%\bibitem{}
Voges,W., Aschenbach,B., Boller,Th., Bräuninger,H., Briel,U.,
Burkert,W., Dennerl,W., Englhauser,J., Gruber,R., Haberl,F.,
Hartner,G., Hasinger,G., K\"urster,M., Pfeffermann,E., Pietsch,W.,
Predehl,P., Rosso,C., Schmitt,J.H.M.M., Tr\"umper,J., Zimmermann,H.-U.,
1996, {\it A\&A}, in press.\\
%\bibitem{}
Weistrop,D., Hintzen,P., Liu,C., Lowenthal,J., Cheng,K.-P.,
Oliversen,R., Brown,L., Woodgate, B., 1995, {\it AJ} {\bf 109}, 981\\
%\bibitem{}
Ugrymov,A., Engels,D., Lipovetsky,V., Hopp,U., Richter,G.,
Izotov,Y.I., Kniazev,A.Y., Popescu,C.C., 1997, {\it Proc. Symp. IAU} 
{\bf 179}, in press \\
%\bibitem{}
Zwicky, F., Herzog,E., Wild,P., 1961-1968, {\it Catalogue of Galaxies and 
Clusters of Galaxies}, Speich, Z\"urich (CGCG).\\
%\end{thebibliography}
}
\end{document}